\documentclass{amsart}

\begin{document}

\newcommand{\nc}{\newcommand}

\nc{\pr}{\noindent{\em Proof. }}
\nc{\g}{\mathfrak g}
\renewcommand{\k}{\mathfrak k}
\nc{\A}{\mathcal A}
\nc{\F}{\mathcal F}
\renewcommand{\H}{\mathfrak H}

\newtheorem{theorem}{Theorem}{}
\newtheorem{lemma}[theorem]{Lemma}{}
\newtheorem{corollary}[theorem]{Corollary}{}
\newtheorem{conjecture}[theorem]{Conjecture}{}
\newtheorem{proposition}[theorem]{Proposition}{}
\newtheorem{axiom}{Axiom}{}
\newtheorem{remark}[theorem]{Remark}{}
\newtheorem{example}{Example}{}
\newtheorem{exercise}{Exercise}{}
\newtheorem{definition}{Definition}{}

\renewcommand{\theremark}{}

\renewcommand{\thedefinition}{\arabic{definition}}

\title{Towards non-perturbative quantization and the mass gap problem for the Yang--Mills Field}

\author[A. Sevostyanov]{A. Sevostyanov}

\address{Institute of Pure and Applied Mathematics,
University of Aberdeen \\ Aberdeen AB24 3UE, United Kingdom \\ e-mail: a.sevastyanov@abdn.ac.uk }


\begin{abstract}
We reduce the problem of quantization of the Yang--Mills field Hamiltonian to a problem for defining a probability measure on an infinite-dimensional space of gauge equivalence classes of connections on $\mathbb{R}^3$. We suggest a formally self-adjoint expression for the quantized Yang--Mills Hamiltonian as an operator on the  corresponding Lebesgue $L^2$-space. In the case when the Yang--Mills field is associated to the abelian group $U(1)$ we define the probability measure which depends on two real parameters $m>0$ and $c\neq 0$. This yields a non-standard quantization of the Hamiltonian of the electromagnetic field, and the associated probability measure is Gaussian. The corresponding quantized Hamiltonian is a self-adjoint operator in a Fock space the spectrum of which is $\{0\}\cup[\frac12m, \infty)$, i.e. it has a gap.
\end{abstract}

\maketitle

\pagestyle{myheadings}

\markboth{A. SEVOSTYANOV}{TOWARDS NON-PERTURBATIVE QUANTIZATION OF THE YANG--MILLS FIELD}

%%%%%%%%%%%%%%%%%%%%%%%%%%%%%%%%%%%%%%%%%%%%%%%%%%%%%%%%%%%%%%%%%%%%%%%%%%%%%%%%%%%%%%%%%%%%%%%%%%%%%%%%%%%%%%%%%%%%%%%%%%%%%%%%%%%%%%%%%%%%%%%%%%%%%%%%%%%%

\section*{Introduction}

\renewcommand{\theequation}{\arabic{equation}}

\setcounter{equation}{0}

The purpose of this short note is to reduce the problem of non-perturbative quantization of the Yang--Mills field Hamiltonian to a problem for defining a probability type measure on an infinite-dimensional space of gauge equivalence classes of connections on $\mathbb{R}^3$. Recall that the Hamiltonian of the Yang--Mills field associated to a compact Lie group $K$ with Lie algebra $\k$ is quadratic in momenta and its potential is equal to the square of the three-dimensional curvature tensor $F$ with respect to a natural metric $\langle \cdot,\cdot\rangle $ on the space of $\k$-valued differential forms on $\mathbb{R}^3$. Our key observation is that the $\k$-valued one-form $G$ on $\mathbb{R}^3$ given by the Hodge star operator $*$ in $\mathbb{R}^3$ applied to $F$, $G=*F$, is a potential vector field on the space of gauge equivalence classes of connections on $\mathbb{R}^3$, the potential being the Chern--Simons functional. So that the potential term of the Yang--Mills Hamiltonian becomes the square of a potential vector field $\langle G,G\rangle $ on the space of gauge equivalence classes of connections on $\mathbb{R}^3$ equipped with the metric $\langle \cdot,\cdot\rangle $ which plays the role of the configuration space of the Yang--Mills field, and the cotangent bundle to it is the corresponding phase space.

We show that for a Riemannian manifold $M$ with a Riemannian metric $\langle \cdot,\cdot\rangle $ any Hamiltonian on the symplectic manifold $T^*M$ of the form 
\begin{equation}\label{classh}
\frac12(\langle p,p\rangle +\langle v(x),v(x)\rangle ),	
\end{equation}
where $p\in T_xM\simeq T^*_xM$ is the momentum and $v={\rm grad}~ \phi$ is a potential vector field, admits a family of canonical quantizations of the form 
\begin{equation}\label{rec}
\frac 12\sum_{a=1}^n\xi_a^*(x)\xi_a(x):L^2(M,d\mu)\rightarrow L^2(M,d\mu).		
\end{equation}
Here $\xi_a(x)$, $a=1,\ldots, {\rm dim}~M$ is an orthonormal basis of $T_xM$, and $\xi_a^*(x)$ is the operator formally adjoint to $\xi_a(x)$ with respect to the canonical scalar product in the space $L^2(M,d\mu)$ of square integrable functions on $M$ with respect to the measure $d\mu=\psi e^{-2\phi}dx$, where $dx$ is the Lebesgue measure on $M$ associated to the Riemannian metric, and $\psi$ is an arbitrary smooth non-vanishing function on $M$.

The appearance of the function $\psi$ shows some ambiguity which is permitted by the correspondence principle in the course of quantization. We shall see that according to this principle for any smooth non-vanishing function $\psi$ on $M$ the operator given by expression (\ref{rec}) is a quantization of the Hamiltonian $\frac12(\langle p,p\rangle +\langle v(x),v(x)\rangle )$. But, of course, the properties of the quantized Hamiltonian depend on the choice of $\psi$. In practice the choice of $\psi$ should be dictated by experimental data and by purely mathematical restrictions. It seems that the freedom of this kind in the quantization of classical Hamiltonian systems has not been used so far. As we shall see the latter type of restrictions becomes primarily important in the case of the Yang--Mills field. 

To illustrate the above mentioned ambiguity we are going to consider the situation when $M=\mathbb{R}$ with the usual Euclidean metric, and the classical Hamiltonian is $\frac12\langle p,p\rangle $, i.e. it describes a free particle on the line. If $\psi(x)=1$ then the corresponding operator (\ref{rec}) is 
$$
-\frac12\frac{d^2}{dx^2}:L^2(\mathbb{R}, dx)\rightarrow L^2(\mathbb{R}, dx),
$$
i.e. it is the quantum Hamiltonian of a free particle on the line. It gives rise to a self-adjoint operator the spectrum of which is $[0,\infty)$.

But one can also choose $\psi(x)=\exp(-\frac12x^2)$, and then the corresponding operator (\ref{rec}) becomes 
$$
-\frac12e^{\frac12x^2}\frac{d}{dx}e^{-\frac12x^2}\frac{d}{dx}:L^2(\mathbb{R},\exp(-\tfrac12x^2)dx)\rightarrow L^2(\mathbb{R},\exp(-\tfrac12x^2)dx)
$$
which is the Hermite differential operator. It gives rise to a self-adjoint operator on $L^2(\mathbb{R},\exp(-\tfrac12x^2)dx)$ the spectrum of which is the set $\{0,\frac12,1,\frac32, \ldots\}$, and the corresponding eigenfunctions are the Hermite polynomials (see e.g. \cite{Herm}). Thus with this choice of $\psi$ we obtain, up to a non-essential constant, the Hamiltonian of a quantum harmonic oscillator, and the spectrum of it has a gap separating it from the zero eigenvalue corresponding to the ground state.

Note that in Quantum Mechanics one has to take $\psi =1$ in the above example in order to make the momentum operator $\frac 1i \frac{d}{dx}$ self-adjoint in $L^2(\mathbb{R}, \psi dx)$. However, in Quantum Field Theory operators of variational derivatives with respect to fields, which play the role of $\frac 1i \frac{d}{dx}$, may have no physical meaning. Therefore they do not need to be self-adjoint, and a non-trivial choice of $\psi$ is allowed.

We show that the Hamiltonian of the Yang--Mills field is of type (\ref{classh}), where $M$ is the space of gauge equivalence classes of connections on $\mathbb{R}^3$ equipped with the metric $\langle \cdot,\cdot\rangle $, and $\phi$ is the Chern--Simons functional which we denote by $CS$. Expressing the corresponding quantized Hamiltonian in form (\ref{rec}) solves the so called normal ordering problem which appears in the course of quantization. Thus the problem of quantization of the Yang--Mills Hamiltonian is reduced to defining a measure on the infinite-dimensional space of gauge equivalence classes of connections on $\mathbb{R}^3$ with ``density'' $\psi e^{-2\phi}$. Note that measures on infinite-dimensional spaces are probability measures, and to ensure that the obtained measure on the space of gauge equivalence classes of connections on $\mathbb{R}^3$ is a probability measure it is natural to choose $\psi=\exp(-\frac12\langle G,G\rangle )$ which guarantees that $\psi e^{-2\phi}$ decreases at ``infinity'' in this space.

It turns out, however, that even in the abelian case when $K=U(1)$ this Ansatz does not work. If we use the Coulomb gauge fixing condition to describe the space of gauge equivalence classes of $U(1)$-connections on $\mathbb{R}^3$ as the space of vector fields satisfying the condition ${\rm div}~A=0$ then the appropriate choice for $\psi$ is  $\exp(-\frac1{2c^2}\langle G,G\rangle -\frac12(c^2+m)\langle A,A\rangle )$, $c,m\in \mathbb{R}$, $c\neq 0$, $m>0$, and we show that
$$
\psi e^{-2\phi}=\exp(-\frac1{2c^2}\langle G,G\rangle -2CS(A)-\frac12(c^2+m)\langle A,A\rangle )	
$$    
is the exponent of a negatively defined quadratic expression in $A$. So that the corresponding probability measure is Gaussian. With this choice of $\psi$ the quantized abelian Yang--Mills field suggested in this paper rather resembles the second harmonic oscillator type quantization of the classical Hamiltonian for a free particle on the line considered above.   

Indeed, we prove that the corresponding quantized Yang--Mills Hamiltonian defined following recipe (\ref{rec}) is self-adjoint and its spectrum is $\{0\}\cup [\frac12m,\infty)$, i.e. it has a gap.

The paper is organized as follows. In Sections \ref{YMH} and \ref{YMPh} we recall the results on the Lagrangian and the Hamiltonian formulation for the Yang--Mills field. These results are well-known in some form. We formulate them in a form suitable for our purposes. In Proposition \ref{YMhamiltprop} we make the key observation about the structure of the potential in the Hamiltonian of the Yang--Mills field.

In Section \ref{quantmod} we discuss quantizations of Hamiltonians of the Yang--Mills type mentioned above, and in Section \ref{YMhamquant} these results are applied to the Yang--Mills Hamiltonian. 

{\bf Acknowledgements}

The results presented in this paper have been partially obtained during research stays at Institut des Haut \'{E}tudes Scientifiques, Paris and Max--Planck--Instut f\"{u}r Mathematik, Bonn. The author is grateful to these institutions for hospitality. 

The research on this project received funding from the European Research Council (ERC) under the European Union's Horizon 2020 research and innovation program (QUASIFT grant agreement 677368) during the visit of the author to Institut des Haut \'{E}tudes Scientifiques, Paris.

The author also thanks the referee for useful comments.

%%%%%%%%%%%%%%%%%%%%%%%%%%%%%%%%%%%%%%%%%%%%%%%%%%%%%%%%%%%%%%%%%%%%%%%%%%%%%%%%%%%%%%%%%%%%%

%\renewcommand{\theequation}{\thesubsection.\arabic{equation}}

%%%%%%%%%%%%%%%%%%%%%%%%%%%%%%%%%%%%%%%%%%%%%%%%%%%%%%%%%%%%%%%%%%%%%%%%%%%%%%%%%%%%%%%%%%%%%%%%%%%%%%%%%%%%%%%%%%%%%%%%%%%%%%%%%%%%%%%%%%%%%%%%%%%%%%%%%%%%%%

%%%%%%%%%%%%%%%%%%%%%%%%%%%%%%%%%%%%%%%%%%%%%%%%%%%%%%%%%%%%%%%%%%%%%%%%%%%%%%%%%%%%%%%%%%%

\section{The Yang--Mills field in Hamiltonian formulation}\label{YMH}

%%%%%%%%%%%%%%%%%%%%%%%%%%%%%%%%%%%%%%%%%%%%%%%%%%%%%%%%%%%%%%%%%%%%%%%%%%%%%%

\subsection{The Yang--Mills field as a Hamiltonian system with constrains}\label{YMh}

In this section following \cite{FS} we recall the Lagrangian and the
Hamiltonian formalism for the Yang--Mills field. The canonical variables and the Hamiltonian
will be obtained via the Legendre transform starting from the Lagrangian
formulation.

Let $K$ be a compact semi-simple Lie group, $\k$ its Lie algebra and $\g$ the
complexification of $\k$. We denote by $(\cdot ,\cdot )$ the Killing form of $\g$.
Recall that the restriction of this form to $\k$ is non-degenerate and negatively defined.
We shall consider the Yang--Mills functional on the affine space of smooth
connections in the trivial $K$-bundle, associated to
the adjoint representation of $K$, over the standard Minkowski space
${\mathbb R}_{1,3}$. Fixing the standard trivialization of this bundle and the
trivial connection as an origin in the affine space of connections we can identify this space
with the space $\Omega^1(\mathbb{R}_{1,3},\k)$ of $\k$-valued 1-forms on ${\mathbb R}_{1,3}$.
Let $\A\in \Omega^1(\mathbb{R}_{1,3},\k)$ be such a connection.
Denote by $\F$ the curvature 2-form of this connection, $\F=d\A + \frac{1}{2} [\A\wedge \A]$.
Here as usual we denote by $[\A\wedge \A]$ the
operation which takes the
exterior product of $\k$-valued 1-forms and the commutator of their values in
$\k$. The Yang--Mills action functional $YM$ evaluated at $\A$ is defined by
the formula

\begin{equation}\label{YM}
YM = \frac 12 \int_{{\mathbb R}_{1,3}} (\F \wedge , \star \F),
\end{equation}
where $\star$ stands for the Hodge star operation associated to the standard
metric on the Minkowski space, and we evaluate the Killing form on the
values of $\F$ and $\star\F$ and also take their exterior product. The
corresponding Lagrangian density $\mathcal{L}$ is equal to $\star(\F \wedge ,
\star\F)$,
\begin{equation}\label{L}
\mathcal{L}=\frac 12 \star (\F \wedge , \star\F).
\end{equation}

Next, following \cite{FS}, we pass from Lagrangian to Hamiltonian
formulation for the Yang--Mills field. To this end one should use the modified
Lagrangian density $\mathcal{L}'$,

\begin{equation}\label{L'}
\mathcal{L}'= \star((d\A + \frac{1}{2} [\A\wedge \A]- \frac 12 \F )\wedge , \star\F),
\end{equation}
where $\A$ and $\F$ should be regarded as independent variables. The
equations of motion obtained by extremizing the corresponding action functional are equivalent
to those derived from the action (\ref{YM}). Indeed, the equation for $\F$
following from (\ref{L'}) is just the definition of the curvature,
\begin{equation}\label{nondyn}
\F=d\A + \frac{1}{2} [\A\wedge \A],
\end{equation}
and the other equation,
$$
d\star \F+[\A \wedge \star \F]=0,
$$
becomes the usual Yang--Mills equation after expressing $\F$
in terms of $\A$.

In order to pass to the Hamiltonian formalism for the Yang--Mills field we
introduce a convenient notation that will be used throughout of this paper.
Let $\Omega^*(\mathbb{R}^{3},\k)$ be the space of
$\k$-valued differential forms on $\mathbb{R}^{3}$. We
define a scalar product on this space, whenever it is finite, by
\begin{equation}\label{prod}
\langle \omega_1,\omega_2\rangle =-\int_{\mathbb{R}^{3}}(\omega_1\wedge,*\omega_2)=
-\int_{\mathbb{R}^{3}}*(\omega_1\wedge,*\omega_2)d^{3}x,~
\omega_{1,2} \in \Omega^*(\mathbb{R}^{3},\k)
\end{equation}
where $*$ stands for the Hodge star operation associated to the standard
Euclidean metric on  $\mathbb{R}^{3}$, and we evaluate the Killing form on the
values of $\omega_1$ and $*\omega_2$ and also take their exterior product.

Let $\A$ be $\k$-valued connection 1-form in the trivial $K$-bundle, associated to
the adjoint representation of $K$, over the standard Minkowski space, $\F$ its curvature
2-form. Fix a coordinate system $(x_0,x_1,x_2,x_3)$ on ${\mathbb R}_{1,3}$ so that $x_0=t$ is the time and $(x_1,x_2,x_3)$ are orthogonal Cartesian coordinates on $\mathbb{R}^{3}\subset {\mathbb R}_{1,3}$. We denote by $A$ the ``three-dimensional Euclidean part'' of $\A$,
$A=\sum_{k=1}^{3}A_kdx_k$, where $A_k=\A_k$ for
$k=1,2,3$. We also introduce the ``electric'' field $E$ and the ``magnetic''
field $G$ associated to $\F$ as follows:
$$
\begin{array}{l}
E=\sum_{k=1}^{3}E_kdx_k,~ E_k=\F_{0k}, \\
\\
G=*F,~ F=dA + \frac{1}{2} [A\wedge A],
\end{array}
$$
i.e. $F$ is the ``three-dimensional'' spatial part of $\F$.

We recall that the covariant derivative
$d_A:\Omega^n(\mathbb{R}^{3},\k)\rightarrow \Omega^{n+1}(\mathbb{R}^{3},\k)$ associated to $A$
is defined by $d_A\omega =d\omega + [A\wedge \omega]$, and the operator
formally adjoint to $d_A$ with respect to scalar product (\ref{prod}) is
equal to $-*d_A*$. We denote by ${\rm div}_A$ the part of this operator acting
from $\Omega^1(\mathbb{R}^{3},\k)$ to $\Omega^{0}(\mathbb{R}^{3},\k)$, with
the opposite sign,
$$
{\rm div}_A=*d_A*:\Omega^1(\mathbb{R}^{3},\k)\rightarrow
\Omega^{0}(\mathbb{R}^{3},\k).
$$

Using this notation the Lagrangian density (\ref{L'}) can be rewritten, up to a divergence, in
the following form
\begin{equation}\label{L''}
\mathcal{L}'=-\left( *(E\wedge,*\partial_t A)-\frac 12 (*(E\wedge,*E)+*(G\wedge,*G))+(\A_0,{\rm div}_A
E)\right).
\end{equation}
It is easier to confirm this formula by an explicit calculation in terms of components (see \cite{IZ}, Section 12-1-4 for further details). Below we use Greek letters to label the coordinates $(x_0,x_1,x_2,x_3)$ in the Minkowski space, Latin letters to label the spacial coordinates $(x_1,x_2,x_3)$, and the usual lifting rules for tensor indexes with the help of metric. $\partial_\mu$ and $\partial_k$ stand for the partial derivative with respect to $x_\mu$ and $x_k$, respectively.
From (\ref{L'}) using the definition of $\star$ we have in terms of the components of the connection and of the curvature forms
$$
\mathcal{L}'= \sum_{\nu<\mu}(\partial_\nu \A_\mu-\partial_\mu \A_\nu+[\A_\nu,\A_\mu]-\frac12\F_{\nu\mu}, \F^{\nu\mu})=
$$
$$
=\sum_{k=1}^3(\partial_t \A_k-\partial_k \A_0+[\A_0,\A_k]-\frac12\F_{0k}, \F^{0k})+
$$
$$
+\sum_{i<j}(\partial_i \A_j-\partial_j \A_i+[\A_i,\A_j]-\frac12\F_{ij}, \F^{ij}).
$$
Now we can rewrite this equation with the help of the non-dynamical equations
$$
\partial_i \A_j-\partial_j \A_i+[\A_i,\A_j]=\F_{ij}
$$
following from (\ref{nondyn}),
$$
\mathcal{L}'
=\sum_{k=1}^3(\partial_t \A_k-\partial_k \A_0+[\A_0,\A_k]-\frac12\F_{0k}, \F^{0k})
+\sum_{i<j}\frac12(\F_{ij}, \F^{ij}).
$$
Recalling the definitions of $F$ and $E$ we arrive at
$$
\mathcal{L}'
=\sum_{k=1}^3\left(-(\partial_t A_k, E_k)+\partial_k (\A_0,E_k)-(\A_0,\partial_k E_k+[A_k,E_k])+\frac12(E_k, E_k)\right)+\sum_{i<j}\frac12(\F_{ij}, \F^{ij})
$$
which is equal to the right hand side of (\ref{L''}) up to the divergence term $\sum_{k=1}^3\partial_k (\A_0,E_k)$.

For the corresponding action we have
\begin{equation}\label{YM''}
 YM'=\int_{-\infty}^{\infty}\left( \langle E,\partial_t A\rangle -\frac 12 (\langle E,E\rangle +\langle G,G\rangle )+\langle \A_0,{\rm div}_A
E\rangle \right) dt.
\end{equation}

Denote $C={\rm div}_A E$, and
introduce an orthonormal basis $T_a,~a=1,\ldots ,{\rm dim}~\k $ in $\k$ with respect to the
Killing form and the components of $A$, $E$, $\A_0$ and $C$ associated to this basis,
$A_k=\sum_a A_k^{a}T_a,~E_k=\sum_a
E_k^{a}T_a,~\A_0=\A_0^{a}T_a,~C=C^{a}T_a$. In terms of these components the
action (\ref{YM''}) takes the form
\begin{equation}\label{YM'''}
YM'=\int_{{\mathbb R}_{1,3}}\left( \sum_{k,a}E_k^{a}\partial_t A_k^{a}-h(A,E)+
\sum_{a}\A_0^{a}C^{a}\right) d^4x,
\end{equation}
where
$$
h(A,E)=-\frac 12 (*(E\wedge,*E)+*(G\wedge,*G))
$$
is the Hamiltonian density. Denote
\begin{equation}\label{hamilt}
H(A,E)=\frac 12 (\langle E,E\rangle +\langle G,G\rangle ).
\end{equation}
From formula (\ref{YM'''}) it is clear that $A_k^{a}$ and $E_k^{a}$ are
canonical conjugate coordinates and momenta for the Yang--Mills field,
$H(A,E)$ is the Hamiltonian, $\A_0^{a}$ are Lagrange multipliers and
$C^a=0$ are constrains imposed on the canonical variables.

The Yang--Mills equations become Hamiltonian with respect to the canonical Poisson
structure
\begin{equation}\label{pois}
  \{E_k^{a}(x),A_l^b(y)\}=\delta_{kl}\delta^{ab}\delta(x-y),
\end{equation}
and all the other Poisson brackets of the components of $E$ and $A$ vanish.
One can also check that
\begin{equation}\label{momentbrack}
\{C^a(x),C^b(y)\}=\sum_c t^{abc}C^c(x)\delta(x-y),
\end{equation}
where $t^{abc}$ are the structure constants of Lie algebra $\k$ with respect
to the basis $T^a$, $[T^a,T^b]=\sum_c t^{abc}T^c$, and that
\begin{equation}\label{haminv}
\{H(A,E), C^a(x)\}=0.
\end{equation}
This means that the Yang--Mills field is a generalized Hamiltonian system
with first class constrains according to Dirac's classification \cite{Dir}.

%%%%%%%%%%%%%%%%%%%%%%%%%%%%%%%%%%%%%%%%%%%%%%%%%%%%%%%%%%%%%%%%%%%%%%%%%%%%%%%%%%%%%%%%%%%

\section{The structure of the phase space of the Yang--Mills field}\label{YMPh}

%%%%%%%%%%%%%%%%%%%%%%%%%%%%%%%%%%%%%%%%%%%%%%%%%%%%%%%%%%%%%%%%%%%%%%%%%%%%%%

\subsection{Reduction of the phase space}\label{YMred}

In this section we collect some facts on the Poisson geometry of the phase
space of the Yang--Mills field and related gauge actions. These results are certainly well
known.
But it seems that they are not presented in the literature in the form suitable for our
purposes (see, however, \cite{Sing} about the gauge actions).

To begin with, we consider the Yang--Mills field as a generalized Hamiltonian system with
Hamiltonian (\ref{hamilt}) and constraints $C={\rm div}_A~E=0$ on the phase space
$\Omega^1_c(\mathbb{R}^{3},\k)\times \Omega^1_c(\mathbb{R}^{3},\k)$ equipped
with Poisson structure (\ref{pois}). Here $\Omega^1_c(\mathbb{R}^{3},\k)$
stands for the space of smooth $\k$-valued 1-forms on $\mathbb{R}^{3}$ with
compact support. Later the phase space will be considerably extended.

The Poisson structure (\ref{pois}) has a natural geometric interpretation.
Indeed, consider the affine space  of smooth connections
in the trivial $K$-bundle, associated to
the adjoint representation of $K$, over ${\mathbb R}^{3}$. As in Section
\ref{YMh} we fix the standard trivialization of this bundle and the
trivial connection as an origin in the affine space of connections and identify this space
with the space $\Omega^1(\mathbb{R}^{3},\k)$ of $\k$-valued 1-forms on ${\mathbb
R}^{3}$. Let $\mathcal{D}$ be the subspace in the affine space of
connections isomorphic to $\Omega^1_c(\mathbb{R}^{3},\k)$ under this
identification. We shall frequently write $\mathcal{D}$ instead of
$\Omega^1_c(\mathbb{R}^{3},\k)$ and call this space the space of compactly
supported $K$-connections on $\mathbb{R}^{3}$.

The space $\mathcal{D}$ has a natural Riemannian metric defined with the help of
scalar product (\ref{prod}),
\begin{equation}\label{riemann}
\langle E,E'\rangle =-\int_{\mathbb{R}^{3}}(E\wedge,*E'),~E,E'\in T_A\mathcal{D}\simeq\mathcal{D},
\end{equation}

This metric gives rise to a natural imbedding $T\mathcal{D}\hookrightarrow
T^*\mathcal{D}$ induced by the natural embeddings
$$
\begin{array}{l}
T_A\mathcal{D}\simeq \mathcal{D}\hookrightarrow \mathcal{D}^*\simeq T_A^*\mathcal{D}, A\in\mathcal{D} \\
\\
\omega \mapsto \hat{\omega},\\
\\
\hat{\omega}(\omega')=\langle \omega,\omega'\rangle ,~\omega,\omega'\in \mathcal{D}.
\end{array}
$$
Using this imbedding the tangent bundle $T\mathcal{D}$ can be
equipped with the natural structure of a Poisson manifold induced by the restriction of
the canonical symplectic structure of $T^*\mathcal{D}$ to $T\mathcal{D}\hookrightarrow
T^*\mathcal{D}$. This restriction is well defined as a symplectic form on $T\mathcal{D}$ since the canonical symplectic form on $T^*\mathcal{D}$ is constant, and metric (\ref{riemann}), with the help of which the restriction of the form to $T\mathcal{D}$ is defined, is non-degenerate. Explicitly this restriction $\Omega$ is given by 
$$
\Omega(A,E)((X,Y),(X',Y'))=\langle Y',X\rangle-\langle Y,X'\rangle,
$$
where $(A,E)\in T\mathcal{D}\simeq \mathcal{D}\times \mathcal{D}$, $(X,Y),(X',Y')\in T_{(A,E)}T\mathcal{D}\simeq \mathcal{D}\times \mathcal{D}$.

The symplectic structure on the space $T\mathcal{D}$ can be identified with that which corresponds to Poisson structure (\ref{pois}).

Now let us discuss the meaning of the constrains. First of all we note that
the constrains $C={\rm div}_A~E$ infinitesimally generate the gauge action on
the phase space $T\mathcal{D}$. More precisely, let $\mathcal{K}$ be the group
of $K$-valued maps $g:\mathbb{R}^{3}\rightarrow K$ such that $g(x)=I$ for
$|x|\geq R(g)$, where $I$ is the identity element of $K$ and $R(g)>0$ is a
real number depending on $g$. $\mathcal{K}$ is called the gauge group of
compactly supported gauge transformations. The Lie algebra of $\mathcal{K}$
is isomorphic to $\Omega^0_c(\mathbb{R}^{3},\k)$.

The gauge group $\mathcal{K}$ acts on the space of connections $\mathcal{D}$
by
\begin{equation}\label{Gaugeact}
\begin{array}{l}
\mathcal{K}\times \mathcal{D} \rightarrow \mathcal{D},\\
\\
g\times A \mapsto g\circ A= -dgg^{-1}+gAg^{-1},
\end{array}
\end{equation}
where we denote $dgg^{-1}=g^*\theta_R$, $gAg^{-1}=\mathrm{Ad}g(A)$, and
$\theta_R$ is the right-invariant Maurer--Cartan form on $K$. This action is free,
so that the quotient $\mathcal{D}/\mathcal{K}$ is a smooth
manifold.

The action (\ref{Gaugeact}) of $\mathcal{K}$ on the space of connections
$\mathcal{D}$ induces an action
\begin{equation}\label{Gaugeactph}
\begin{array}{l}
\mathcal{K}\times T\mathcal{D}\rightarrow T\mathcal{D}, \\
\\
g\times (A,E) \mapsto (gEg^{-1},-dgg^{-1}+gAg^{-1}),
\end{array}
\end{equation}
where as before we write $gEg^{-1}=\mathrm{Ad}g(E)$. This action gives rise
to an action of the Lie algebra $\Omega^0_c(\mathbb{R}^{3},\k)$ of the gauge
group $\mathcal{K}$ on $T\mathcal{D}$ by vector fields. If $X\in \Omega^0_c(\mathbb{R}^{3},\k)$
then the corresponding vector field $V_X(A,E)$ is given by
\begin{equation}\label{gaugeactph}
V_X(A,E)=([X,E],-dX+[X,A]),~(A,E)\in T\mathcal{D}\simeq \mathcal{D}\times \mathcal{D}.
\end{equation}
Here we, of course, identify
$T_{(A,E)}T\mathcal{D}\simeq T\mathcal{D}\simeq\mathcal{D}\times \mathcal{D}$.

The action (\ref{gaugeactph}) is generated by the constraint ${\rm div}_A~E$ in the sense that
for $X\in \Omega^0_c(\mathbb{R}^{3},\k),~(A,E)\in T\mathcal{D}$ we have
$$
\{\langle {\rm div}_A~E,X\rangle ,A(x)\}=-dX(x)+[X(x),A(x)],
$$
and
$$
\{\langle {\rm div}_A~E,X\rangle ,E(x)\}=[X(x),E(x)].
$$

Using the language of Poisson geometry and taking into account formula (\ref{momentbrack})
for the Poisson brackets of the constrains one can say that
$\mathcal{K}\times T\mathcal{D}\rightarrow T\mathcal{D}$ is a Hamiltonian
group action, and the map
\begin{equation}\label{YMmoment}
\begin{array}{l}
\mu:T\mathcal{D}\rightarrow \Omega^0_c(\mathbb{R}^{3},\k),\\
\\
\mu(A,E)={\rm div}_A~E
\end{array}
\end{equation}
is the moment map for this action. In particular, action (\ref{Gaugeactph})
preserves the symplectic form of $T\mathcal{D}$.

We note that action
(\ref{Gaugeactph}) also preserves Riemannian structure (\ref{riemann}) of the configuration
space $\mathcal{D}$.
This follows from the fact that the Killing form on $\k$ is
invariant with respect to the adjoint action of $K$.

The properties of the phase space of the Yang--Mills field
and of the gauge action discussed above are formulated in the following proposition.
\begin{proposition}\label{YMprop}
Let $\mathcal{D}$ be the space of compactly supported $K$-connections on
$\mathbb{R}^{3}$, $\mathcal{K}$ the group of compactly supported gauge
transformations. Then the following statements are true.

(i) The space $\mathcal{D}$ is an infinite dimensional Riemannian
manifold equipped with the metric
\begin{equation}\label{r1}
\langle E,E'\rangle =-\int_{\mathbb{R}^{3}}(E\wedge,*E'),~E,E'\in T_A\mathcal{D}.
\end{equation}
This metric induces a natural imbedding $T\mathcal{D}\hookrightarrow
T^*\mathcal{D}$, and the tangent bundle $T\mathcal{D}$ can be
equipped with the natural structure of a Poisson manifold induced by
the canonical symplectic structure of $T^*\mathcal{D}$.

(ii) The gauge action $\mathcal{K}\times \mathcal{D}\rightarrow \mathcal{D}$
preserves Riemannian metric (\ref{r1}) and
gives rise to a Hamiltonian group action
$\mathcal{K}\times T\mathcal{D}\rightarrow T\mathcal{D}$ with the moment map
$$
\begin{array}{l}
\mu:T\mathcal{D}\rightarrow \Omega^0_c(\mathbb{R}^{3},\k),\\
\\
\mu(A,E)={\rm div}_A~E,~(A,E)\in T\mathcal{D}\simeq
\mathcal{D}\times\mathcal{D}.
\end{array}
$$

(iii) The action of the gauge group $\mathcal{K}$ on the spaces $\mathcal{D}$ and $T\mathcal{D}$
is free, and the reduced phase space $\mu^{-1}(0)/\mathcal{K}$ is a smooth manifold.
\end{proposition}

Finally we make a few remarks on the structure of the Hamiltonian of the
Yang--Mills field.

Since the Hamiltonian $H(A,E)$ of the Yang--Mills field is invariant under
the gauge action (\ref{Gaugeactph}) (this fact can be checked directly and
also follows from formula (\ref{haminv})) the generalized Hamiltonian
dynamics in the sense of Dirac (see \cite{Dir}) described by this Hamiltonian together with the constrains ${\rm div}_A~E=0$ is equivalent, in the sense explained in \cite {FS}, Sect. 3.2, to the usual one on the reduced phase space $\mu^{-1}(0)/\mathcal{K}$ (see \cite{A}, Appendix 5). More explicitly, since $H(A,E)$ is gauge invariant the Hamiltonian vector field of it is tangent to $\mu^{-1}(0)$ and is invariant under the action of $\mathcal{K}$. Thus this vector field gives rise to a Hamiltonian vector field on $\mu^{-1}(0)/\mathcal{K}$ (see \cite{A}, Appendix 5C) which gives rise to a Hamiltonian dynamics on the reduced space.

The Hamiltonian (\ref{hamilt}) itself has a very standard structure; $H(A,E)$ is
equal to the sum of a half of the square of the momentum, $\frac 12
\langle E,E\rangle $, and of a potential $U(A)$, $U(A)=\frac 12 \langle G,G\rangle $. The potential $U(A)$ is, in
turn, equal to a half of the square of the vector field $G\in
\Gamma(T\mathcal{D})$. By definition the vector field $G$ is invariant with respect
to the gauge action of $\mathcal{K}$, $G(g\circ A)=gG(A)g^{-1}$.
The value of this field at each point $A\in \mathcal{D}$
belongs to the kernel of the operator ${\rm div}_A$.
Indeed, from the Bianchi identity
$d_AF=0$, the definition
of $G=*F$ and the formula $**={\rm id}$ it follows that
$$
{\rm div}_A~G=-*d_A**F=-*d_A~F=0.
$$

The vector field $G$ has one more important property: it is potential with the potential function equal to the Chern--Simons functional. Recall that this functional is defined by
\begin{equation}\label{CS}
CS(A)=\frac{1}{2}\langle A,*dA\rangle +\frac{1}{3}\langle A,*[A\wedge A]\rangle .	
\end{equation}

This functional is invariant under the action of the Lie algebra of the gauge group and its gradient is equal to $G$. Note that the Chern--Simons functional is not invariant under the action of the gauge group itself: there is a constant $\kappa$ such that for any $g \in \mathcal{K}$ $CS(g\circ A)$ differs from $CS(A)$ by $\kappa n$, where $n\in \mathbb{Z}$ depends on the homotopy class of $g$ (see e.g. \cite{F}).

Now we summarize the properties of the Hamiltonian of the Yang--Mills field.
\begin{proposition}\label{YMhamiltprop}
(i) The generalized Hamiltonian system on the Poisson manifold $T\mathcal{D}$ with the
Hamiltonian $H(A,E)$,
$H(A,E)=\frac 12 (\langle E,E\rangle +\langle G,G\rangle )$, $G=*F$, $F=dA+\frac{1}{2}[A\wedge A]$, and the constrains
${\rm div}_A~E=0$
describes the Yang--Mills dynamics on $T\mathcal{D}$.

(ii) The Hamiltonian $H(A,E)$ is invariant under the gauge action
$\mathcal{K}\times T\mathcal{D}\rightarrow T\mathcal{D}$ and the generalized Hamiltonian
dynamics described by this Hamiltonian together with the constrains ${\rm div}_A~E=0$
is equivalent to the usual one on the reduced phase space
$\mu^{-1}(0)/\mathcal{K}$.

(iii) The vector field $G$ is invariant with respect
to the gauge action of $\mathcal{K}$, $G(g\circ A)=gG(A)g^{-1}$.
The value of this field at each point $A\in \mathcal{D}$
belongs to the kernel of the operator ${\rm div}_A$,
$$
G(A)\in {\rm Ker}~{\rm div}_A~\forall A\in \mathcal{D}.
$$

(iv) The vector field $G$ is potential with the potential equal to the Chern--Simons functional (\ref{CS}) which is invariant under the action of the Lie algebra of the gauge group.
\end{proposition}

%%%%%%%%%%%%%%%%%%%%%%%%%%%%%%%%%%%%%%%%%%%%%%%%%%%%%%%%%%%%%%%%%%%%%%%%%%%%%%%%%%%%%%%%%%%%%%%%

\subsection{The structure of the reduced phase space}\label{redcoord}

In Propositions \ref{YMprop} and \ref{YMhamiltprop} we formulated all the properties of the Yang--Mills field which are important for our further consideration. In this section we study an arbitrary Hamiltonian system satisfying these properties.

First we consider a phase space equipped with a Lie group action of the type
described in Proposition \ref{YMprop}. Actually
the Riemannian metric introduced in that proposition is only important for the definition of the Hamiltonian of the Yang--Mills field. This metric is not relevant to Poisson geometry. We used this metric in the description of the phase space in order to avoid analytic difficulties arising in the infinite-dimensional case. Now let us forget about the metric for a moment and discuss the geometry of the reduced space.

The Poisson structure described in Proposition \ref{YMprop} is an example of
the canonical Poisson structure on the cotangent bundle, and the group action
on this bundle is induced by a group action on the base manifold.
Thus we start with a manifold $\mathcal{M}$ and a Lie group
$G$ freely acting on $\mathcal{M}$. The canonical symplectic structure on $T^*\mathcal{M}$
can be defined as follows (see \cite{A}).

Denote by $\pi:T^*\mathcal{M}\rightarrow \mathcal{M}$ the
canonical projection, and define a 1-form $\theta$ on $T^*\mathcal{M}$ by
$\theta (v)=p(d\pi v)$,
where $p\in T_x^*\mathcal{M}$ and $v\in T_{(x,p)}(T^*\mathcal{M})$. Then the canonical
symplectic form on $T^*\mathcal{M}$ is equal to $d\theta$.

Recall that the induced Lie group action
$G\times T^*\mathcal{M}\rightarrow T^*\mathcal{M}$ is a Hamiltonian group action
with a moment map $\mu:T^*\mathcal{M}\rightarrow \g^*$, where $\g^*$ is the
dual space to the Lie algebra $\g$ of $G$. The moment map $\mu$ is
uniquely determined by the formula (see \cite{Per}, Theorem 1.5.2)
\begin{equation}\label{mom}
(\mu(x,p),X)=\theta ({\widetilde{X}})(x,p)=p(\widehat X(x)),
\end{equation}
where $\widehat{X}$ is the vector field on
$\mathcal{M}$ generated by arbitrary element $X\in \g$, ${\widetilde X}$ is the
induced vector field on $T^*\mathcal{M}$ and $(,)$ stands for the
canonical paring between $\g$ and $\g^*$.

Formula (\ref{mom}) implies that for
any $x\in \mathcal{M}$ the map $\mu(x,p)$ is linear in $p$. We denote this linear map by $m(x)$,
$m(x):T_x^*\mathcal{M}\rightarrow \g^*$,
\begin{equation}\label{m}
m(x)p=\mu(x,p).
\end{equation}

Next, following \cite{A}, Appendix 5, with some modifications of the proofs suitable for
our purposes, we describe the structure of the reduced space
$\mu^{-1}(0)/{G}$. We start with a simple lemma.
\begin{lemma}\label{l1}
The annihilator $T_x\mathcal{O}^\perp\in T_x^*\mathcal{M}$ of the tangent
space $T_x\mathcal{O}$ to the $G$-orbit $\mathcal{O}\subset \mathcal{M}$ at point $x$ is
isomorphic to ${\rm Ker}~m(x)$, $T_x\mathcal{O}^\perp={\rm Ker}~m(x)$.
\end{lemma}
\pr
First we note that the space $T_x\mathcal{O}^\perp$ is spanned by the
differentials of $G$-invariant functions on $\mathcal{M}$. But from the
definitions of the moment map and of the Poisson structure on
$T^*\mathcal{M}$ we have
\begin{equation}\label{to}
L_{\widehat{X}}f(x)=\{(X,\mu),f\}(x)=(X,m(x)df(x)),
\end{equation}
where $\widehat{X}$ is the vector field on $\mathcal{M}$ generated by
element $X\in \g$, $f\in C^{\infty}(\mathcal{M})$, and $(,)$ stands for the
canonical paring between $\g$ and $\g^*$.

Formula (\ref{to}) implies that $f$ is $G$-invariant if and only if
$df(x)\in {\rm Ker}~m(x)$. This completes the proof.

\qed

\begin{proposition}\label{redstruct}
The action of the group $G$ on $T^*\mathcal{M}$
is free, and the
reduced phase space $\mu^{-1}(0)/{G}$ is a smooth manifold. Moreover, we have
an isomorphism of symplectic manifolds, $\mu^{-1}(0)/{G}\simeq
T^*(\mathcal{M}/{G})$, where $T^*(\mathcal{M}/{G})$ is equipped with the
canonical symplectic structure.
Under this isomorphism
$T_{\mathcal{O}_x}^*(\mathcal{M}/{G})\simeq T_x\mathcal{O}_x^\perp$, where
$\mathcal{O}_x$ is the $G$-orbit of $x$.
\end{proposition}

\pr
Let $\mathcal{O}_x$ be the  $G$-orbit of point $x\in \mathcal{M}$ and
$\pi:\mathcal{M}\rightarrow \mathcal{M}/{G}$ the canonical projection,
$\pi(x)=\mathcal{O}_x$. Denote by $\Xi$ the foliation of the space $\mathcal{M}$ by the
subspaces $T_x\mathcal{O}^\perp$. Since the foliation $\Xi$ is $G$-invariant
and
${\rm Ker}~d\pi|_{T_x\mathcal{M}}=T_x\mathcal{O}_x$ we can identify the
subspace $T_x\mathcal{O}_x^\perp$ with the tangent space
$T_{\mathcal{O}_x}^*(\mathcal{M}/{G})$ by means of the dual map to
the differential of the projection
$\pi$. But the definition of the moment map $\mu$ and Lemma \ref{l1}
imply that
$\mu^{-1}(0)=\{(x,p)\in T^*\mathcal{M}:p \in T_x\mathcal{O}_x^\perp\}$. Therefore
the quotient $\mu^{-1}(0)/{G}$ is diffeomorphic to
$T^*(\mathcal{M}/{G})$, the diffeomorphism being induced by the canonical
projection $\pi$.

From the definitions of
the Poisson structures on $T^*(\mathcal{M}/{G})$ and on the reduced space
$\mu^{-1}(0)/{G}$ it follows that the diffeomorphism
$\mu^{-1}(0)/{G}\simeq T^*(\mathcal{M}/{G})$ is actually an isomorphism of symplectic
manifolds.

\qed

Using the last proposition one can easily describe the space $\Gamma T^*(\mathcal{M}/{G})$ of
covector fields on $\mathcal{M}/{G}$.
\begin{corollary}\label{vectred}
The space $\Gamma T^*(\mathcal{M}/{G})$ is isomorphic to the space of
$G$-invariant sections $V\in \Gamma T^*\mathcal{M}$ such that $V(x)\in
T_x\mathcal{O}_x^\perp$ for any $x\in \mathcal{M}$. Such covector fields
will be called vertical $G$-invariant covector fields on $\mathcal{M}$. We
denote this space by $\Gamma_G^\perp T^*\mathcal{M}$,
$\Gamma_G^\perp T^*\mathcal{M}\simeq \Gamma T^*(\mathcal{M}/{G})$.
\end{corollary}

Now we discuss the class of Hamiltonians on $T^*\mathcal{M}$ we are
interested in. First, recalling Proposition \ref{YMprop} we equip the manifold
$\mathcal{M}$ with a Riemannian metric $\langle \cdot ,\cdot \rangle $ and
assume that the action of $G$ on $\mathcal{M}$  preserves
this metric. Using this metric we can establish an isomorphism of
$G$-manifolds, $T\mathcal{M}\simeq T^*\mathcal{M}$.
We shall always identify the tangent and the cotangent bundle of
$\mathcal{M}$ and the spaces of vector and covector fields on $\mathcal{M}$
by means of this isomorphism.
The tangent bundle $T\mathcal{M}$
will be regarded as a symplectic manifold with the induced symplectic structure.
Under the identification $T\mathcal{M}\simeq T^*\mathcal{M}$ the subspace
$T_x \mathcal{O}^\perp\subset T_x^*\mathcal{M}$ is isomorphic to the
orthogonal complement of the tangent space $T_x\mathcal{O}$ in
$T_x\mathcal{M}$. Note also that since
$T_{\mathcal{O}_x}^*(\mathcal{M}/{G})\simeq T_x\mathcal{O}_x^\perp$
and the metric on $\mathcal{M}$ is $G$-invariant
$T_{\mathcal{O}_x}^*(\mathcal{M}/{G})$ has a scalar product induced
from $T_x\mathcal{O}_x^\perp$, i.e. $\mathcal{M}/{G}$ naturally
becomes a Riemannian manifold. We shall also identify
$T^*(\mathcal{M}/{G})\simeq T(\mathcal{M}/{G})$ by means of the metric.
Denote by $\Gamma_G^\perp T\mathcal{M}$ the space of $G$-invariant vertical vector fields on
$\mathcal{M}$. By Corollary \ref{vectred} we have an isomorphism,
$\Gamma_G^\perp T\mathcal{M}\simeq \Gamma T(\mathcal{M}/{G})$.

On the symplectic manifold $T\mathcal{M}$ we define a Hamiltonian of the
type described in Proposition \ref{YMhamiltprop}. In order to do that we fix
a $G$-invariant vertical vector field $V$ on $\mathcal{M}$. Then we put
$$
l(x,p)=\frac 12 (\langle p,p\rangle +\langle V(x),V(x)\rangle ),~p\in T_x\mathcal{M}.
$$
This Hamiltonian is obviously $G$-invariant and gives rise to a Hamiltonian
$l_{red}$ on the reduced space $\mu^{-1}(0)/{G}\simeq T^*(\mathcal{M}/{G})$.
Since by Corollary \ref{vectred} $V$ can be regarded as a (co)vector field on $\mathcal{M}/{G}$
we have
\begin{equation}\label{hamiltred}
l_{red}(\mathcal{O}_x,p_\perp)=\frac 12
(\langle p_\perp,p_\perp\rangle +\langle V(x),V(x)\rangle ),~p_\perp \in T_x\mathcal{O}_x^\perp\simeq
T_{\mathcal{O}_x}^*(\mathcal{M}/{G}).
\end{equation}

Now we can apply the above obtained results in the case of the Yang--Mills field. 
The reduced phase space of the Yang--Mills field is of the type considered in Lemma \ref{l1} and Proposition \ref{redstruct} with $\mathcal{M}=\mathcal{D}$ and $G=\mathcal{K}$. In the infinite-dimensional case we have to distinguish between $T\mathcal{D}$ and $T^*\mathcal{D}$. But according to Proposition \ref{YMprop} for the description of the Yang--Mills dynamics it suffices to consider $T\mathcal{D}$ and equip it with the Poisson structure induced by the imbedding $T\mathcal{D}\subset T^*\mathcal{D}$ with the help of metric (\ref{r1}). Then the action of $\mathcal{K}$ of $T\mathcal{D}$ becomes Hamiltonian, and in the notation of Lemma \ref{l1} $m(x)={\rm div}_A$.

Let $\mathcal{O}_A$ be the gauge orbit of a connection $A\in \mathcal{D}$. By Lemma \ref{l1} the space $T_{\mathcal{O}_A}\mathcal{D}/\mathcal{K}$ is isomorphic to the kernel of the operator ${\rm div}_A$ in $T_A\mathcal{D}$. The metric (\ref{r1}) induces a Riemannian metric on $\mathcal{D}/\mathcal{K}$ which we denote by the same symbol.

According to Proposition \ref{YMhamiltprop} the vector field $G$ on the space $\mathcal{D}$ is $\mathcal{K}$-invariant and horizontal. Hamiltonian (\ref{hamilt}) is of type (\ref{hamiltred}). Therefore from formula (\ref{hamiltred}) and Proposition \ref{redstruct} we infer that Hamiltonian (\ref{hamilt}) gives rise to the Hamiltonian 
\begin{equation}\label{Hamiltred}
H_{red}(\mathcal{O}_A,E_\perp)=\frac 12
(\langle E_\perp,E_\perp\rangle +\langle G,G\rangle ),~E_\perp \in T_{\mathcal{O}_A}(\mathcal{D}/\mathcal{K})\simeq {\rm Ker~div}_A
\end{equation}
on the reduced phase space $\mu^{-1}(0)/\mathcal{K}\simeq T\mathcal{D}/\mathcal{K}$.

Based on the results of this section we can also make two remarks on the
structure of the gauge orbit space $\mathcal{D}/\mathcal{K}$.
\begin{remark}
The Riemannian geometry of the space $\mathcal{D}/\mathcal{K}$ is
nontrivial. In particular, its curvature tensor is not identically equal to
zero (see \cite{Sing}). This is the main peculiarity of non-abelian gauge
theories.
\end{remark}

\begin{remark}
The quotient $\mathcal{D}/\mathcal{K}$ cannot be
realized as a cross-section for the gauge action of $\mathcal{K}$ on
$\mathcal{D}$.
For any local cross-section of this action there are $\mathcal{K}$-orbits in
$\mathcal{D}$ which
meet this cross-section many times. This phenomenon is called the Gribov
ambiguity (see \cite{Sing1}).

The Riemannian manifold $\mathcal{D}/\mathcal{K}$ cannot be
realized as a cross-section for the action of $\mathcal{K}$ on
$\mathcal{D}$ even locally.
This is due to the fact that the
foliation $\Xi$ of $\mathcal{D}$ by the subspaces ${\rm Ker~div}_A\subset T_A\mathcal{D}$
is not an integrable distribution, and therefore the
subspaces ${\rm Ker~div}_A$ are not tangent to a submanifold in
$\mathcal{D}$. Indeed, the components $C^a$ of the constraint ${\rm div}_A~E$ regarded as an
$\Omega^0(\mathbb{R}^3,\k)$-valued 1-form on $\mathcal{D}$ do not form a
differential ideal. Therefore the conditions of the Frobenius integrability theorem are
not satisfied. In Poisson geometry constrains of this type are called
non-holonomic.
\end{remark}

%%%%%%%%%%%%%%%%%%%%%%%%%%%%%%%%%%%%%%%%%%%%%%%%%%%%%%%%%%%%%%%%%%%%%%%%%%%%%%%%%%%%%%%%%%%%%%%
%%%%%%%%%%%%%%%%%%%%%%%%%%%%%%%%%%%%%%%%%%%%%%%%%%%%%%%%%%%%%%%%%%%%%%%%%%%%%%%%%%%%%%%%%%%%%%%%%

\section{Quantization of the Hamiltonian of the Yang--Mills field}

%%%%%%%%%%%%%%%%%%%%%%%%%%%%%%%%%%%%%%%%%%%%%%%%%%%%%%%%%%%%%%%%%%%%%%%%%%%%%%%%%%%%%%%%%%%%%%%%

\subsection{Quantization of Yang--Mills type Hamiltonians: a model case}\label{quantmod}

Let $M$ be $n$-dimensional Riemannian manifold with a metric $\langle \cdot,\cdot\rangle $. For simplicity we denote the pairing between $T_xM$ and $T^*_xM$ and the induced scalar product on $T_x^*M$ by the same symbol as the metric on $M$. As before we can identify $T^*M$ and $TM$ using the metric.

Consider a Hamiltonian of type (\ref{hamiltred}) on $T^*M\simeq TM$,
\begin{equation}\label{hamM}
h(x,p)=\frac 12 (\langle p,p\rangle +\langle v(x),v(x)\rangle ),~x\in M,~p\in T_x M,	
\end{equation}
where $v$ is a vector field on $M$. So $M$ plays the role of $\mathcal{M}/G$ in this section. 

Assume that the vector field $v$ is potential with a potential function $\phi$, so $v={\rm grad}~\phi$.

Let $\xi_a(x)$, $a=1,\ldots, n$ be an orthonormal basis in $T_x M$, $\langle \xi_a,\xi_b\rangle =\delta_{ab}$.  
Let $T_{\xi_a}=\langle \xi_a,p\rangle -i\langle \xi_a,v\rangle $, $T_{\xi_a}^*=\langle \xi_a,p\rangle +i\langle \xi_a,v\rangle $. From this definition and from the definition of the basis $\xi_a$ it follows immediately that
\begin{equation}
h(x,p)=\frac12\sum_{a=1}^nT_{\xi_a}^*T_{\xi_a}.	
\end{equation}

Now let $x^1,\ldots ,x^n$ be a local coordinate system on $M$ defined on an open subset of $M$, $\xi_a^i$ the coordinates of $\xi_a$ with respect to this coordinate system, so $\xi_a=\sum_{i=1}^n\xi_a^i\frac{\partial}{\partial x^i}$. %Then one verifies that on this subset $\sum_{a=1}^n\xi_a^i(x)\xi_a^j(x)=g^{ij}(x)$, where $g^{ij}(x)$ is the induced metric on $T^*M$ in terms of the coordinates $x^1,\ldots ,x^n$. 
Denote by $g_{ij}$ the components of metric tensor of the metric $\langle \cdot,\cdot\rangle $ in terms of the coordinates $x^1,\ldots ,x^n$.  We also have $p=\sum_{i=1}^np_idx^i$.

Let $L^2(M,\psi)$ be the Hilbert space of complex-valued functions on $M$ such that
$$
\int_M|f|^2\psi d\mu<\infty,
$$
where $\mu$ is the Lebesgue measure on $M$ associated to the Riemannian metric, and $\psi\in C^\infty(M)$ is a smooth non-vanishing function on $M$. The scalar product on $L^2(M,\psi)$ is given by the usual formula 
$$
(f,f')_\psi=\int_M f\bar{f}'\psi d\mu.
$$

According to the canonical quantization philosophy and the correspondence principle after quantization $p_i$ becomes the  operator $\frac1i\frac{\partial}{\partial x^i}$ in $L^2(M,\psi)$, and any function of $x$ becomes the multiplication operator by that function in $L^2(M,\psi)$, so $T_{\xi_a}$ becomes the operator $\frac1i\xi_a-i(\xi_a,v)=-i\nabla_{\xi_a}$, where $\nabla_{\xi_a}=\xi_a+\langle \xi_a,v\rangle $. 

We would like to define a self-adjoint operator in $L^2(M,\psi)$ which is a quantization of the Hamiltonian $h(x,p)$. According to the canonical quantization philosophy we have to ensure that the quantized Hamiltonian becomes a self-adjoint operator in $L^2(M,\psi)$. In order to fulfill this requirement we have to require that after quantization $T_{\xi_a}^*$ becomes the operator adjoint to $\frac1i\xi_a-i(\xi_a,v)$ in $L^2(M,\psi)$. In terms of the local coordinates the operator formally adjoint to $\frac1i\xi_a-i\langle \xi_a,v\rangle $ takes the form 
$$
f\mapsto i\left( -\frac{1}{\sqrt{g}}\psi^{-1}\frac{\partial}{\partial x^i}(\xi_a^i\sqrt{g}\psi f)+\langle \xi_a,v\rangle f\right)=i\nabla_{\xi_a}^*f,
$$
where $g=|{\rm det}~g_{ij}|$, so a natural candidate for a quantized Hamiltonian is the self-adjoint operator $h_0$ defined by the expression
\begin{equation}\label{qh1}
\frac 12\sum_{a=1}^n\nabla_{\xi_a}^*\nabla_{\xi_a}.	
\end{equation}

One straightforwardly verifies that, after applying reversely the correspondence principle according to which the  operator $\frac1i\frac{\partial}{\partial x^i}$ becomes $p_i$, and the multiplication operator by a function in $L^2(M,\psi)$ becomes this function in the classical limit, expression (\ref{qh1}) becomes Hamiltonian (\ref{hamM}) in the classical limit.

Note that the operator of multiplication by $e^{\phi}$ gives rise to a unitary equivalence $L^2(M,\psi)\rightarrow L^2(M,\psi e^{-2\phi})$, and the operator $h$ in $L^2(M,\psi e^{-2\phi})$ unitarily equivalent to $h_0$, $h=e^{\phi}h_0e^{-\phi}$, is defined using the expression 
\begin{equation}\label{expr}
e^{\phi}\frac 12\sum_{a=1}^n\nabla_{\xi_a}^*\nabla_{\xi_a}e^{-\phi}=\frac 12\sum_{a=1}^n\xi_a^*\xi_a,	
\end{equation}
where as above in local coordinates $\xi_a=\sum_{i=1}^n\xi_a^i\frac{\partial}{\partial x^i}$, and $\xi_a^*$ is the operator formally adjoint to $\xi_a$ with respect to the scalar product in $L^2(M,\psi e^{-2\phi})$.
  
A formal definition of the self-adjoint operator $h$ can be given using its bilinear form. Clearly, expression(\ref{expr}) defines a non-negative symmetric operator on $L^2(M,\psi e^{-2\phi})$, with the domain being the space $C_0^\infty$ of smooth complex-valued compactly supported functions on $M$. Thus one can apply the Friedrichs extension method to define its self-adjoint extension (see \cite{RS2}, Theorem X.23). This yields the following statement.
\begin{theorem}
The non-negative bilinear form $(f,f')_h=\frac 12\sum_{a=1}^n(\xi_af,\xi_a f')_{\psi e^{-2\phi}}$, with the domain being the space $C_0^\infty$ of smooth complex-valued compactly supported functions on $M$, is closable on $L^2(M,\psi e^{-2\phi})$ with a domain $D$ and its closure defines a non-negative self-adjoint operator $h$ on $L^2(M,\psi e^{-2\phi})$ with a domain $D(h)$, so that $(f,f')_h=(hf,f')_{\psi e^{-2\phi}}$ for any $f\in D(h), f'\in D$.
 
Moreover, if the constant function $1$ belongs to $L^2(M,\psi e^{-2\phi})$ then $1$ is an eigenfunction of the operator $h$ with the lowest eigenvalue zero.

For any smooth non-vanishing function $\psi$ on $M$, $\psi\in C^\infty(M)$, the operator $h$ is a quantization of Hamiltonian (\ref{hamM}) in the sense of canonical quantization. 
\end{theorem}

The second part of the previous theorem ensures the existence of the lowest energy ground state for the operator $h$.

%%%%%%%%%%%%%%%%%%%%%%%%%%%%%%%%%%%%%%%%%%%%%%%%%%%%%%%%%%%%%%%%%%%%%%%%%%%%%%%%%%%%%%%%%%%%%%%

\subsection{Application to the Yang--Mills Hamiltonian}\label{YMhamquant}

Now we are going to apply the idea of the previous section to quantize the reduced Yang--Mills Hamiltonian defined by formula (\ref{Hamiltred}) on the reduced phase space $\mu^{-1}(0)/\mathcal{K}\simeq T\mathcal{D}/\mathcal{K}$. Note that according to  this formula $H_{red}$ is of the same type as the Hamiltonian $h(x,p)$ considered in the previous section with $\phi=CS(A)$. So informally, according to Proposition \ref{redstruct}, we should take $\mathcal{D}/\mathcal{K}$ as $M$ in the previous section. But the fact that $\mathcal{D}/\mathcal{K}$ is infinite-dimensional now brings further difficulties.

According to the philosophy of Section \ref{quantmod}, firstly we should try to find a measure with ``density'' which resembles $\psi e^{-2\phi}$ with $\phi=CS(A)$ and an appropriate $\psi$. The peculiarity of the infinite-dimensional case is that the existence of such measures is a very strong condition. In particular, all known measures of this kind are probability measures, so that the entire space has a finite volume usually normalized to one. Therefore $\psi e^{-2\phi}$ should rapidly decrease at infinity. As it can be easily seen this condition is not fulfilled if we choose $\psi=1$. It is natural to use $\psi=\exp(-\frac12\langle G,G\rangle )$ and then 
\begin{equation}\label{ansatz}
\psi e^{-2\phi}=\exp(-\frac12\langle G,G\rangle -2CS(A)).	
\end{equation}
This functional is invariant under the action of the Lie algebra of the gauge group. The Chern--Simons functional is not invariant under the action of the gauge group itself, so functional (\ref{ansatz}) is not quite well defined on $\mathcal{D}/\mathcal{K}$. But one should not expect that a measure on an infinite-dimensional quotient space by an action of an infinite-dimensional group is induced by a measure on the original space invariant under the group action. This phenomenon is related to the fact that there are no even translation invariant measures on infinite-dimensional spaces. Therefore firstly we have to fix a model for $\mathcal{D}/\mathcal{K}$ and then define a measure on it. Thus we only need to define a functional on the model for $\mathcal{D}/\mathcal{K}$ the gradient of which coincides with that of $CS(A)$. This is the only condition required by the correspondence principle. Note that the gradient of $CS(A)$ is well defined as a vector field on $\mathcal{D}/\mathcal{K}$.  

Even taking into account the discussion above it turns out that a measure with a ``density'' which resembles $\exp(-\frac12\langle G,G\rangle -2CS(A))$ still does not exist even in the abelian case, and a certain ``renormalization'' is required to define it. We shall construct this measure now in the abelian case when $K=U(1)$.

So from now on we assume that $K=U(1)$. We identify the corresponding Lie algebra with $\mathbb{R}$. Choose a model $\mathcal{D}_0$ for $\mathcal{D}/\mathcal{K}$ being the space of the elements $A$ of $\mathcal{D}$ which satisfy the condition ${\rm div}A=0$, where ${\rm div}={\rm div}_0$. Note that in the abelian case $CS(A)$ is gauge invariant and gives rise to a functional on $\mathcal{D}/\mathcal{K}$. Its restriction to $\mathcal{D}_0$ will be denoted by the same letter. All functionals of $A$ below will be considered as functionals on $\mathcal{D}_0$.  
 
In the abelian case we have 
\begin{equation}\label{exp1}
\exp(-\frac12\langle G,G\rangle -2CS(A))=\exp(-\frac12\langle *dA,*dA\rangle -\langle *dA,A\rangle ), 	
\end{equation}
and this function is the exponent of an expression which is quadratic in $A$ which means that the measure that we are going to construct is likely to be Gaussian. To define such a measure we have to ensure that the expression in the exponent is negative definite which is not true for (\ref{exp1}). In order to fulfill this condition we choose $\psi=\exp(-\frac1{2c^2}\langle G,G\rangle -\frac12(c^2+m)\langle A,A\rangle )$, $A\in \mathcal{D}_0$, where $c,m\in \mathbb{R}$ are constants, $c\neq 0$, and $m>0$. Then
\begin{equation}\label{exp2}
\psi e^{-2\phi}=\exp(-\frac1{2c^2}\langle G,G\rangle -2CS(A)-\frac12(c^2+m)\langle A,A\rangle )=	
\end{equation}
$$
=\exp(-\frac1{2c^2}\langle *dA,*dA\rangle -\langle *dA,A\rangle -\frac12(c^2+m)\langle A,A\rangle )=\exp(-\frac12(\Lambda A,A)),
$$
where $\Lambda=T^2+m{\rm Id}$, and $T=\frac1c{\rm curl}+c{\rm Id}$, ${\rm curl}=*d$ are symmetric operators on $\mathcal{D}_0$ with respect to the scalar product $\langle \cdot,\cdot\rangle $.

Recall that Gaussian measures are actually defined on spaces dual to nuclear spaces (see e.g. \cite{Ob}). This forces us to enlarge $\mathcal{D}_0$ and to replace it with the nuclear space $\mathcal{S}_0$ which consists of elements $A$ of $\Omega^1(\mathbb{R}^3,\mathfrak{k})=\Omega^1(\mathbb{R}^3)$ the components of which with respect to the fixed Cartesian coordinate system belong to the Schwartz space and which satisfy the condition ${\rm div}A=0$, the topology on $\mathcal{S}_0$ being induced by that of the Schwartz space. Let $\mathcal{S}_0^*$ be the dual space. 

According to the Bochner--Minlos theorem (Theorem 1.5.2 in \cite{Ob}) Gaussian measures on $\mathcal{S}_0^*$ are Fourier transforms of characteristic functionals on $\mathcal{S}_0$, and the Gaussian measure with ``density'' which resembles $\exp(-\frac12(\Lambda A,A))$ should have the characteristic functional $C(A)=\exp(-\frac12\langle \Lambda^{-1} A,A\rangle )$.

\begin{lemma}\label{charf}
$C(A)$, $A\in \mathcal{S}_0$ is a characteristic functional, i.e. 
\begin{enumerate}
\item
$C(A)$, $A\in \mathcal{S}_0$ is positive definite: for any $\alpha_1,\ldots, \alpha_n\in \mathbb{C}$, $\xi_1,\ldots, \xi_n\in \mathcal{S}_0$ we have $\sum_{i,j=1}^n\alpha_i\bar{\alpha}_jC(\xi_i-\xi_j)\geq 0$;
\item
$C(A)$, $A\in \mathcal{S}_0$ is a continuous functional on $\mathcal{S}_0$; 
\item
$C(0)=1$.
\end{enumerate}
\end{lemma} 

To justify this claim we shall need some facts about the spectral decomposition for the operator curl (see \cite{BS3}, \S 8.6, Ex. 4). 

Let $\mathcal{H}^i$, $i=0,1$ be the completion of the space $\Omega^i_c(\mathbb{R}^3,\k)=\Omega^i_c(\mathbb{R}^3)$ with respect to scalar product (\ref{prod}). Here we assume that $\k$ is identified with $\mathbb{R}$ and the Killing form is just minus the product of real numbers. According to Lemma 8 (i) in \cite{SS} ${\rm div}: \Omega^1_c(\mathbb{R}^3)\rightarrow \mathcal{H}^0$ is a closable operator. We denote its closure by the same symbol, ${\rm div}: \mathcal{H}^1\rightarrow \mathcal{H}^0$. 

${\rm Ker~div}\subset \mathcal{H}^1$ is naturally a Hilbert space with the scalar product inherited from $\mathcal{H}^1$, and, in fact, this Hilbert space is rigged. Namely,
\begin{equation}\label{GG}
\mathcal{S}_0\subset {\rm Ker~div}\subset \mathcal{S}_0^*
\end{equation}
is the corresponding Gelfand--Graev triple.

Let $\mathcal{H}^i_{\mathbb{C}}$, $i=0,1$, ${\rm Ker~div}_{\mathbb{C}}$ and $\mathcal{S}_0^{\mathbb{C}}$ be the complexifications of $\mathcal{H}^i$, $i=0,1$, ${\rm Ker~div}$ and $\mathcal{S}_0$, respectively.  We can identify $\mathcal{H}^1_{\mathbb{C}}$ with the Lebesgue space $L^2(\mathbb{R}^3,\mathbb{C}^3)$ of square integrable functions with values in $\mathbb{C}^3$ equipped with the scalar product induced from $\mathcal{H}^1_{\mathbb{C}}$. The componentwise Fourier transform $F$ provides an isomorphism of $L^2(\mathbb{R}^3,\mathbb{C}^3)$ onto itself under which ${\rm Ker~div}_{\mathbb{C}}$ is mapped onto the subspace $F({\rm Ker~div}_{\mathbb{C}})$ in $L^2(\mathbb{R}^3,\mathbb{C}^3)$ which consists of $\mathbb{C}^3$-valued functions $f(k)\in L^2(\mathbb{R}^3,\mathbb{C}^3)$, $k\in \mathbb{R}^3$ satisfying the condition $k\cdot f(k)=0$, where $\cdot$ is the standard scalar product in $\mathbb{C}^3$ induced by the Cartesian product in $\mathbb{R}^3$ fixed above. Also the Fourier transform maps $\mathcal{S}_0^{\mathbb{C}}\subset \mathcal{H}^1_{\mathbb{C}}$ isomorphically onto the subspace $F(\mathcal{S}_0^{\mathbb{C}})$ of $\mathbb{C}^3$-valued functions $f(k)\in L^2(\mathbb{R}^3,\mathbb{C}^3)$, $k\in \mathbb{R}^3$ with components from the complex Schwartz space and satisfying the condition $k\cdot f(k)=0$. ${\rm Ker~div}_{\mathbb{C}}$ is an invariant subspace for the natural extension of the operator curl to $\mathcal{H}^1_{\mathbb{C}}$. Note that the action of curl on $\mathcal{H}^1_{\mathbb{C}}$ preserves $\mathcal{H}^1$ and ${\rm Ker~div}$, i.e. curl is a real operator and ${\rm Ker~div}$ is an invariant subspace for it. ${\rm curl}: {\rm Ker~div}_{\mathbb{C}}\rightarrow {\rm Ker~div}_{\mathbb{C}}$ is a self-adjoint operator with the natural domain $\{v\in {\rm Ker~div}_{\mathbb{C}}:{\rm curl}v\in {\rm Ker~div}_{\mathbb{C}}\}$.

Under the isomorphism $F$ the operator curl acting on $\mathcal{H}^1_{\mathbb{C}}\simeq L^2(\mathbb{R}^3,\mathbb{C}^3)$ becomes the operator $F{\rm curl}F^{-1}$ acting by the cross vector product by $ik$ on elements of $f(k)\in L^2(\mathbb{R}^3,\mathbb{C}^3)$. For each $k\in \mathbb{R}^3$ this operator acts of $f(k)\in \mathbb{C}^3$ by the matrix 
\begin{equation}\label{cr}
\left(\begin{array}{ccc}
0 & -ik_3 & ik_2 \\
ik_3 & 0 & -ik_1 \\
-ik_2 & ik_1 & 0
\end{array}
\right)
\end{equation}
which is nothing but the symbol of the operator curl. 

$F({\rm Ker~div}_{\mathbb{C}})$ is an invariant subspace for the operator $F{\rm curl}F^{-1}$. For each fixed $k$ the eigenvalues of matrix (\ref{cr}) restricted to the subspace in $\mathbb{C}^3$ which consists of elements $v\in \mathbb{C}^3$ satisfying the condition $k\cdot v=0$ are $\pm |k|$. 
According to \cite{BS3}, \S 8.6, Ex. 4 this implies that the spectrum of the operator curl is absolutely continuous, $\sigma({\rm curl})=\sigma_{ac}({\rm curl})=\mathbb{R}$,
and hence curl has no eigenvectors in the usual sense. But it has a complete basis of generalized eigenvectors (see, for instance, \cite{GV}).

Namely, this operator can be easily diagonalized by means of
the Fourier transform (see \cite{BS3}, Ch. 8, \S 8.6, Ex. 4). The generalized complex eigenvectors corresponding to the generalized
eigenvalues $\pm |k|$, $k\in \mathbb{R}^3\setminus \{ 0 \}$, can be chosen, for instance, in the form
\begin{equation}\label{roteigen}
\frac{1}{\sqrt{2}(2\pi)^{\frac{3}{2}}}e^{ik\cdot x}(\theta_1(k)\pm
i\theta_2(k)),
\end{equation}
where $\theta_{1,2}(k)$ are 1-forms on $\mathbb{R}^3$ dual to orthonormal vectors
$e_{1,2}(k)$, with respect to the fixed Cartesian scalar product, such that for
every $k\neq 0$ $\frac{k}{|k|},e_1(k),e_2(k)$ is an orthonormal basis in
$\mathbb{R}^{3}$, $\frac{k}{|k|}\times e_1(k)=e_2(k)$ (vector product) and
$e_{1,2}(k)$ smoothly depend on $k\in \mathbb{R}^{3}\setminus \{ 0\}$.

Since the operator ${\rm curl}$ sends real-valued functions to real valued functions one can also find real generalized eigenvectors $e_{\pm}(k)\in \mathcal{S}_0^*$ corresponding to $\pm |k|$, $k\in \mathbb{R}^3\setminus \{ 0 \}$.

The vectors $e_{\pm}(k)$ are generalized eigenvectors for the operator $\mathrm{curl}$
in the sense that
\begin{equation}\label{eigen}
\langle e_{\pm}(k),\mathrm{curl}~\omega\rangle =\pm|k|\langle e_{\pm}(k),\omega\rangle ~\mathrm{for~any}~\omega \in \mathcal{S}_0.
\end{equation}
Note also that $\mathcal{S}_0$ is dense in $\mathrm{Ker}~\mathrm{div}$.

{\em Proof of Lemma \ref{charf}.}
Firstly we show that $\Lambda^{-1}:\mathcal{S}_0\rightarrow \mathcal{S}_0$ is a continuous operator. The easiest way to see this is to observe that according to the results on the eigenvalues of matrix (\ref{cr}) mentioned above the eigenvalues of the symbol of the operator $\Lambda$ acting for each fixed $k$ on the subspace in $\mathbb{C}^3$ which consists of elements $v\in \mathbb{C}^3$ satisfying the condition $k\cdot v=0$ are $(\pm\frac1c|k|+c)^2+m\geq m>0$. Therefore the inverse to the symbol is well-defined for each $k$, the entries of the inverse to the symbol are smooth and bounded (in fact they are rational functions of the components of $k$ with respect to the orthonormal basis in $\mathbb{R}^3$), and hence it gives rise to a bounded operator on $F(\mathcal{S}_0^{\mathbb{C}})$. Applying the inverse to the Fourier transform and recalling that $\Lambda$, and hence $\Lambda^{-1}$, preserve $\mathcal{S}_0\subset \mathcal{S}_0^{\mathbb{C}}$ we deduce that $\Lambda^{-1}:\mathcal{S}_0\rightarrow \mathcal{S}_0$ is a continuous operator.  

Recall also that $\langle \cdot,\cdot\rangle $ is a continuous bilinear form on $\mathcal{S}_0$. Therefore the functional $C(A)=\exp(-\frac12\langle \Lambda^{-1} A,A\rangle )$ is continuous. Obviously, $C(0)=1$.

Finally we have to check that $C(A)$ is positive definite.
Note that  $\langle \Lambda \cdot,\cdot\rangle =\langle T\cdot,T\cdot\rangle +m\langle \cdot,\cdot\rangle $. Therefore  $\langle \Lambda \cdot,\cdot\rangle $ is a positive definite bilinear form on $\mathcal{S}_0$, and $\Lambda$ is a positive operator on ${\rm Ker~div}$. Thus $\Lambda^{-1}$ is a positive operator on ${\rm Ker~div}$ as well, and $\langle \Lambda^{-1} \cdot,\cdot\rangle $ is a positive definite bilinear form on $\mathcal{S}_0$. 

Note that the results on the spectrum of the operator curl above imply that the spectrum of the real operator $\Lambda$ acting on the complexification of ${\rm Ker~div}$ is continuous and coincides with the set $[m,\infty)$, and therefore the spectrum of the real operator $\Lambda^{-1}$ acting on the complexification of ${\rm Ker~div}$ is continuous and coincides with the set $[0,\frac1m]$.
We deduce that the bilinear form $\langle \Lambda^{-1} \cdot,\cdot\rangle $ on $\mathcal{S}_0$ is non-degenerate and defines the structure of a pre-Hilbert space on $\mathcal{S}_0$.

Now by Lemma 2.1.1 in \cite{Ob} $C(A)$ is positive definite, i.e. for any $\alpha_1,\ldots, \alpha_n\in \mathbb{C}$, $\xi_1,\ldots, \xi_n\in \mathcal{S}_0$ we have $\sum_{i,j=1}^n\alpha_i\bar{\alpha}_jC(\xi_i-\xi_j)\geq 0$.

Thus $C(A)$, $A\in \mathcal{S}_0$ is a positive definite continuous functional on $\mathcal{S}_0$ satisfying $C(0)=1$, i.e. it is a characteristic functional.
 
\qed

Using the Bochner--Minlos theorem (Theorem 1.5.2 in \cite{Ob}) we immediately deduce the following corollary of Lemma \ref{charf}.
\begin{corollary}
There is a probability measure $\mu$ on $\mathcal{S}_0^*$ such that $C(A)$ is the Fourier transform of $\mu$, i.e.
$$  
C(A)=\int_{\mathcal{S}_0^*}e^{i\langle x,A\rangle }d\mu(x),
$$
where $\langle x,A\rangle $ stands for the pairing of $x\in \mathcal{S}_0^*$ and $A\in \mathcal{S}_0$.
\end{corollary}

Let $\mathcal{H}=L^2(\mathcal{S}_0^*,\mu)$ be the usual complex Lebesgue space associated to the measure $\mu$. Let $\xi_a$, $a\in \mathbb{N}$ be an orthonormal basis of ${\rm Ker~div}$ which consists of elements from $\mathcal{S}_0$. 
For any $\xi\in \mathcal{S}_0^*$ denote by $D_\xi$ the Gateaux derivative for functions on $\mathcal{S}_0^*$, i.e. for $F:\mathcal{S}_0^*\rightarrow \mathbb{C}$
$$
D_\xi F(x)=\frac{d}{dt}|_{t=0}F(x+t\xi).
$$
Let $D^*_\xi$ be the operator formally adjoint to $D_\xi$ in $\mathcal{H}$.

Any $\xi\in \mathcal{S}_0$ defines a linear function on $\mathcal{S}_0^*$, $X_\xi(x)=\langle x,\xi\rangle $. Let $\mathcal{P}$ be the algebra of functions on $\mathcal{S}_0^*$ generated by complex polynomials in variables $X_\xi$, $\xi\in \mathcal{S}_0$.

According to the philosophy developed in the previous section (see formula (\ref{expr})) the operator defined by the expression
\begin{equation}\label{YMexp}
H=\frac12\sum_{a=1}^\infty D^*_{\xi_a}D_{\xi_a}	
\end{equation}
can be regarded as a quantization of the Yang--Mills Hamiltonian $H_{red}$. Here each $\xi_a\in \mathcal{S}_0$ is regarded as an element of $\mathcal{S}_0^*$ via imbedding (\ref{GG}).

\begin{proposition}
Expression (\ref{YMexp}) does not depend on the choice of the basis $\xi_a$, $a\in \mathbb{N}$ and defines an operator in $\mathcal{H}$ with domain $\mathcal{P}$ which is essentially self-adjoint. We denote its self-adjoint closure by the same letter. The self-adjoint operator $H:\mathcal{H}\rightarrow \mathcal{H}$ defined this way can be regarded as a quantization of the Yang--Mills Hamiltonian $H_{red}$.
\end{proposition}

\pr
Similarly to the discussion in \cite{Hid}, Ch. 11, page 408, one can see that expression (\ref{YMexp}) does not depend on the choice of the basis $\xi_a$, $a\in \mathbb{N}$ and using the arguments from the proof of Theorem 11.1 in \cite{Hid} verbatim one can immediately deduce that this expression defines an operator in $\mathcal{H}$ with domain $\mathcal{P}$ which is essentially self-adjoint. 

\qed

The spectral decomposition for the operator $H$ can be performed in the usual way using a Fock space presentation for $\mathcal{H}$ which can be constructed as follows. 

Recall that the generalized Fourier transform
$$
\Phi:\mathrm{Ker}~\mathrm{div}\rightarrow
L^2_{+}(\mathbb{R}^{3})\stackrel{\cdot}{+}
L^2_{-}(\mathbb{R}^{3}),
$$
where $L^2_{\pm}(\mathbb{R}^{3})$
are copies of the usual $L^2(\mathbb{R}^{3})$, associated to
the basis $e_{\pm}(k)$ of generalized eigenvectors is given in terms of components by
\begin{equation}\label{fouriero}
\Phi(\omega)_{\pm}(k)=-L^{2}\text{-}\lim_{R\rightarrow \infty}\int_{|x|\leq R}
*(\omega\wedge *{e_{\pm}(k)})d^3x, \Phi(\omega)_{\pm}\in L^2_{\pm}(\mathbb{R}^{3}), \Phi(\omega)=\Phi(\omega)_+\stackrel{\cdot}{+}\Phi(\omega)_-.
\end{equation}
Here $L^{2}\text{-}\lim$ stands for the limit with respect to the
$L^2(\mathbb{R}^{3})$-norm.

For $\omega \in \mathcal{S}_0$ we can also
write
\begin{equation}\label{fouriero1}
\begin{array}{l}
\Phi(\omega)_{\pm}(k)=-\lim_{R\rightarrow \infty}\int_{|x|\leq R}
*(\omega\wedge *{e_{\pm}(k)})d^3x= \\
\\
=-\lim_{R\rightarrow \infty}\int_{\mathbb{R}^3}
*(\omega\wedge *{\chi_R(x)e_{\pm}(k)})d^3x=\langle \omega,e_{\pm}(k)\rangle ,
\end{array}
\end{equation}
where $\chi_R(x)$ is the characteristic function of the ball of radius $R$.

Since the usual Fourier transform is unitary, one can normalize the generalized eigenvectors $e_{\pm}(k)$ in such a way that $\Phi$ is also a unitary map. We shall always assume that such normalization is fixed.

Using the generalized eigenvectors $e_{\pm}(k)$ and unitarity of $\Phi$ operator (\ref{YMexp}) can be rewritten in the following form
\begin{equation}\label{YMsym}
H=\frac12\sum_{\varepsilon=\pm}\int_{\mathbb{R}^3}d^3k D^*_{e_{\varepsilon}(k)}D_{e_{\varepsilon}(k)}.
\end{equation}

Note that by Proposition 4.3.11 in \cite{Ob} for $\xi,\eta\in \mathcal{S}_0\subset \mathcal{S}_0^*$ the operators $D_\xi,D_\eta^*$ satisfy the following commutation relations
\begin{equation}\label{commrel}
[D_\xi,D^*_\eta]=(\Lambda\xi,\eta),[D_\xi,D_\eta]=[D_\xi^*,D^*_\eta]=0.	
\end{equation}

By definition the operator $\Lambda$ acts on the generalized eigenvectors $e_{\varepsilon}(k)$ as follows $\Lambda e_{\varepsilon}(k)=((\frac1c\varepsilon|k|+c)^2+m)e_{\varepsilon}(k)$.

The last two observations imply that one can establish a Hilbert space isomorphism between $\mathcal{H}$ and a Fock space as follows (see \cite{Ber}, Ch. 1, \cite{Ob}, Theorem 2.3.5).

Let $H_{\varepsilon_1,\ldots, \varepsilon_n}$ be the space of complex-valued symmetric functions $f(k_1,\ldots, k_n)$ of $n$ variables $k_i\in \mathbb{R}^3$, $i=1,\ldots, n$ such that 
$$
\left\|f\right\|_{\varepsilon_1,\ldots, \varepsilon_n}^2=\int_{(\mathbb{R}^3)^n}|f(k_1,\ldots, k_n)|^2\prod_{i=1}^n((\varepsilon_i\frac1c|k_i|+c)^2+m)dk_1\ldots dk_n<\infty.
$$

$H_{\varepsilon_1,\ldots, \varepsilon_n}$ is a Hilbert space with the scalar product
$$
(f,g)_{\varepsilon_1,\ldots, \varepsilon_n}=\int_{(\mathbb{R}^3)^n}f(k_1,\ldots, k_n)\overline{g(k_1,\ldots, k_n)}\prod_{i=1}^n((\varepsilon_i\frac1c|k_i|+c)^2+m)dk_1\ldots dk_n.
$$

Let $\mathcal{F}$ be the space of all sequences $(f_{\varepsilon_1,\ldots, \varepsilon_n})_{n=0}^\infty$, $f_{\varepsilon_1,\ldots, \varepsilon_n}\in H_{\varepsilon_1,\ldots, \varepsilon_n}$ such that
$$
\sum_{n=0}^\infty\sum_{\varepsilon_1,\ldots,\varepsilon_n=\pm}\left\|f_{\varepsilon_1,\ldots, \varepsilon_n}\right\|_{\varepsilon_1,\ldots, \varepsilon_n}^2<\infty,
$$
where we assume that for $n=0$ $H_{\varepsilon_1,\ldots, \varepsilon_n}=\mathbb{C}$ with the usual complex number scalar product and norm.

$\mathcal{F}$ is a Hilbert space with the scalar product induced from
$$\bigoplus_{n=0}^\infty\bigoplus_{\varepsilon_1,\ldots,\varepsilon_n=\pm} H_{\varepsilon_1,\ldots, \varepsilon_n}.$$ 

The following result is the standard Wiener--It\^{o}--Segal isomorphism between $\mathcal{F}$ and $\mathcal{H}$ (see \cite{Ob}, Theorem 2.3.5, \cite{Ber}, Ch. 1).
\begin{theorem}\label{FTR}
The map
$\mathcal{F}\rightarrow \mathcal{H}$,
$$
(f_{\varepsilon_1,\ldots, \varepsilon_n})_{n=0}^\infty\mapsto\sum_{n=0}^\infty\sum_{\varepsilon_1,\ldots,\varepsilon_n=\pm}\int_{(\mathbb{R}^3)^n}dk_1\ldots dk_nf_{\varepsilon_1,\ldots, \varepsilon_n}D_{e_{\varepsilon_1}(k_1)}^*\ldots D^*_{e_{\varepsilon_n}(k_n)}1
$$
is a well-defined Hilbert space isomorphism.
\end{theorem}

$D_{e_{\varepsilon_1}(k_1)}^*\ldots D^*_{e_{\varepsilon_n}(k_n)}1$ can be regarded as elements of a space of generalized functionals on $\mathcal{S}_0^*$. We are not going to define this space here (see e.g. \cite{Hid}, Ch. 3, 4). 

Commutation relations (\ref{commrel}), formula (\ref{YMsym}) for $H$ and the unitarity of the generalized Fourier transform $\Phi$ imply that the elements $D_{e_{\varepsilon_1}(k_1)}^*\ldots D^*_{e_{\varepsilon_n}(k_n)}1$ and the constant function $1$ are the generalized eigenvectors of the operator $H$. Namely, at least formally, we have
\begin{equation}\label{geneigen}
HD_{e_{\varepsilon_1}(k_1)}^*\ldots D^*_{e_{\varepsilon_n}(k_n)}1=\frac12\sum_{i=1}^n((\varepsilon_i\frac1c|k_i|+c)^2+m)D_{e_{\varepsilon_1}(k_1)}^*\ldots D^*_{e_{\varepsilon_n}(k_n)}1,	
\end{equation}
$$
H1=0.
$$
By Theorem \ref{FTR} the set of the generalized eigenvectors is complete. Therefore using the formulas for the generalized eigenvalues in (\ref{geneigen}) we deduce the following statement.
\begin{theorem}
The spectrum of the operator $H$ is $\{0\}\cup[\frac12m,\infty)$. The eigenspace corresponding to the eigenvalue $0$ is one-dimensional and is generated by the constant function $1\in \mathcal{H}=L^2(\mathcal{S}_0^*,\mu)$ which can be regarded as the ground state. The other points of the spectrum belong to the absolutely continuous spectrum which is of Lebesgue type. The spectral  multiplicity function takes the constant value $\mathbb{N}$ on the absolutely continuous spectrum. Thus $\sigma_{pt}(H)=\{0\}$, $\sigma_{ac}(H)=[\frac12m,\infty)$, $\sigma(H)=\sigma_{pt}(H)\cup\sigma_{ac}(H)$, and the spectrum of $H$ has a gap. 
\end{theorem}

\begin{remark}
Note that the condition $m>0$ is essential in the above construction of the quantization of the abelian Yang--Mills field. The standard quantization of the abelian Yang--Mills field used in Quantum Electrodynamics yields a massless theory. In contrast to our quantization the quantized Hamiltonian in Quantum Electrodynamics cannot be realized as a self-adjoint operator in an $L^2$-space. This quantization is unlikely to have a counterpart in the non-abelian case and looks rather exceptional. 
\end{remark}
 
In conclusion we remark that in the non-abelian case a properly quantized Hamiltonian $H_{red}$ should act as a self-adjoint operator in an $L^2$-space associated to a measure with a ``density'' which resembles functional (\ref{ansatz}) with an appropriate ``renormalization''. If this measure was constructed the quantized Hamiltonian would be immediately defined.

\end{document}